\documentclass[a4paper,10pt]{article}
\usepackage[margin=1.3cm]{geometry}
\usepackage{comment}
\usepackage{amssymb,extarrows,graphicx,subfigure,setspace}
\usepackage{cite}
\usepackage{slashed}
\usepackage{tensor}
\usepackage[toc,page]{appendix}
\usepackage{color}
\usepackage{physics}
\usepackage{hyperref}
\usepackage{dirtytalk}
\hypersetup{colorlinks=true, linkcolor=blue, citecolor=red, linktoc=page}
\makeatother

\usepackage{float}
\usepackage{textcomp}
\usepackage{amsmath}
\newmuskip\pFqmuskip

\newcommand*\pFq[6][8]{%
  \begingroup 
  \pFqmuskip=#1mu\relax
  \mathcode`\,=\string"8000
  \begingroup\lccode`\~=`\,
  \lowercase{\endgroup\let~}\pFqcomma
  {}_{#2}F_{#3}{\left[\genfrac..{0pt}{}{#4}{#5};#6\right]}%
  \endgroup
}
\newcommand{\pFqcomma}{\mskip\pFqmuskip}

\usepackage{mathrsfs}
\usepackage{soul}
\usepackage{hyperref}

\newcommand{\be}{\begin{equation}}
\newcommand{\bea}{\begin{eqnarray}}
\newcommand{\eea}{\end{eqnarray}}
\newcommand{\ba}{\begin{array}}
\newcommand{\ea}{\end{array}}
\newcommand{\ee}{\end{equation}}
\newcommand{\bes}{\begin{equation*}}
\newcommand{\beas}{\begin{eqnarray*}}
\newcommand{\eeas}{\end{eqnarray*}}
\newcommand{\bas}{\begin{array*}}
\newcommand{\eas}{\end{array*}}
\newcommand{\ees}{\end{equation*}}

\numberwithin{equation}{section}

\begin{document}
\color{black}
\begin{center}
\Large{\bf Cooling and heating regions of Joule-Thomson expansion for AdS black holes: Einstein-Maxwell-Power-Yang-Mills and Kerr Sen black holes}\\
\small \vspace{0.7cm}

\small {\bf Mohammad Reza Alipour$^{\dag,\star}$\footnote {Email:~~~mr.alipour@stu.umz.ac.ir}}, \quad
\small {\bf Saeed Noori Gashti $^{\dag}$\footnote {Email:~~~saeed.noorigashti@stu.umz.ac.ir; saeed.noorigashti70@gmail.com}},\quad
\small {\bf Mohammad Ali S. Afshar $^{\star,\ddagger}$\footnote {Email:~~~m.a.s.afshar@gmail.com}},\quad
\small {\bf Jafar Sadeghi$^{\star,\ddagger}$\footnote {Email:~~~pouriya@ipm.ir}}, \quad \\
\vspace{0.5cm}$^{\star}${Department of Physics, Faculty of Basic
Sciences, University of Mazandaran\\ P. O. Box 47416-95447, Babolsar, Iran}\\
\vspace{0.3cm}$^{\dag}${School of Physics, Damghan University, P. O. Box 3671641167, Damghan, Iran}\\
\vspace{0.3cm}$^{\ddagger}${Canadian Quantum Research Center, 204-3002 32 Ave Vernon, BC V1T 2L7, Canada}
\small \vspace{1cm}
\end{center}
\begin{abstract}
In this paper, we study the Joule-Thomson Expansion (JTE) process for two types of black holes: AdS-Einstein–Maxwell-Power-Yang–Mills (AEMPYM) and AdS-Kerr-Sen (AKS). Our study focuses on understanding how various parameters influence the Joule-Thomson Coefficient (JTC), the inversion curve, and the ratio of minimum inversion temperature to critical temperature. For the AKS black hole, we observe that the isenthalpic curves can exhibit either cooling or heating behavior. This behavior is determined by the inversion curve, which is affected by the black hole's mass and specific parameters such as $b$ (parameter signifies the ionic charge of the black hole) and $a$ (rotation parameter). In the case of the AEMPYM black hole, our findings reveal that the ratio of minimum inversion temperature to critical temperature approaches a specific value as Maxwell's charge increases. This ratio remains constant for certain parameter values, while it varies for others. Specifically, when the parameter $q$ (real positive parameter of AEMPYM black hole) is greater than 1, the ratio is almost equal to 1/2 as Maxwell's charge (C) increases. When q equals 1/2, the ratio is exactly 1/2 for all values of (C). For values of (q) between 1/2 and 1, the ratio is close to 1/2, and for values of (q) between 0 and 1/2, the ratio decreases, moving away from 1/2. For the AKS black hole, we find that specific parameter values, such as (a = 0.00951) and (b = 0.00475 ), yield a ratio of minimum inversion temperature to a critical temperature that is approximately 1/2. This consistency across different parameter values highlights the robustness of our findings. Finally, we compare our results with those reported in the existing literature, providing a comprehensive summary in detailed tables. This comparison not only validates our findings but also situates them within the broader context of black hole thermodynamics and the Joule-Thomson effect.\\\\
Keywords; Joule-Thomson expansion, AdS-Maxwell-power-Yang-Mills black holes, AdS-Kerr-Sen black holes\\\\
Corresponding author; Saeed Noori Gashti
\end{abstract}
\tableofcontents
\section{Introduction}
The insights provided by astronomy, astrophysics, and experimental cosmology suggest that cosmic structures adhere to the same principles and foundations observed on Earth, albeit with minor variations and necessary simplifications. This structural and fractal similarity is a crucial tool for human creativity and cognition, enabling us to decode more aspects of this global pattern daily. Researchers leverage this understanding in hypothesizing theoretical models, which may not yet be empirically verifiable\cite{600}. They construct various models based on mathematical logic and physical laws, exploring all potential scenarios to predict the most plausible ideas consistent with the formation of the mysterious, infinite universe. The existing black hole models are one of the most obvious examples of this form of inference\cite{601}. Despite being one of the most important cosmic entities, they are studied based on this pattern. The comparison of a black hole's gravitational behavior with a thermodynamic ensemble in the 1970s gave rise to a significant branch of black hole physics, namely black hole thermodynamics. For instance, the four laws of black hole mechanics bear a striking resemblance to the laws of thermodynamics \cite{1,2}.
Similarly, a black hole's surface area is analogous to entropy in thermodynamics, and its surface gravity is comparable to temperature\cite{1,2}. The introduction of Maldacena duality, also known as the AdS/CFT correspondence, established a deeper connection between black hole thermodynamics and this duality, offering profound insights into quantum gravity. In the context of AdS/CFT, a black hole's entropy in the bulk AdS space correlates with the entropy of the corresponding CFT on the boundary, known as the Bekenstein-Hawking entropy \cite{3}. Moreover, the black hole's temperature is related to the CFT's temperature\cite{3,4,5,6}.\\\\
This correspondence provides a powerful tool to study the quantum aspects of gravity and black holes using the methods of quantum field theory. In AdS space, there is a Hawking-Page phase transition between a stable large black hole and a thermal gas\cite{7}. This phase transition is a first-order transition that occurs when the temperature of the system reaches a critical value, where the free energy of the black hole becomes lower than that of the thermal gas\cite{7}.
This phase transition can be interpreted as a confinement/deconfinement phase transition of a gauge field\cite{8}.
The Hawking-Page phase transition can be seen as a transition from a deconfined phase in the thermal gas to a confined phase in the black hole\cite{8}. When the AdS black holes have an electric charge, they exhibit rich phase structures that were studied by Chamblin et al\cite{9,10}. They found that the phase transition behavior of charged AdS black holes resembles the liquid-gas phase transition in a van der Waals system\cite{11}. In the extended phase space where the cosmological constant is treated as pressure\cite{12}, the Pressure-Volume (P-V) critical behavior of charged AdS black holes was investigated and it was shown that they have a similar analogy to the van der Waals liquid-gas system. In addition to the phase transition and critical phenomena\cite{13,14,15,16,17,18,19,20}, the analogy between the black holes and the van der Waals system was also creatively applied to the well-known JTE process\cite{21} recently. This means that the JTC can be used to study the thermodynamics of black holes as well\cite{21',21''}. For example, the isenthalpic expansion process is the analog of the JTE process for black holes, where the black hole mass is constant while the black hole pressure and volume are changed. The black hole pressure is related to the cosmological constant, and the black hole volume is related to the horizon radius. In this case, the inversion curve is the curve that separates the regions where the black hole temperature increases or decreases as the pressure decreases. One of the intriguing features of the inversion curves for black hole systems is that they have  positive slopes, unlike the van der Waals system, which has both positive and negative slopes. For example, for charged AdS black holes \cite{21}  and Kerr-AdS black holes \cite{48}, the isenthalpic expansion process and the inversion curve have been analyzed and observed that the inversion curve was found to have a positive slope, meaning that the black hole temperature always decreases as the pressure decreases. Then the analysis was generalized to other types of AdS black holes, such as quintessence charged AdS black holes\cite{23}, holographic superfluids \cite{24}, charged AdS black holes in f(R) gravity \cite{25}, AdS black holes with a global monopole \cite{26}, and AdS black holes in Lovelock gravity \cite{27}. For further study, you can see in \cite{28,29,30,31,32,33,34,35,36,36.1,37}.\\ In general, the study of the thermodynamics of black holes in different frameworks, such as extended, CFT, and RPS phase space, as well as the study of phase transitions, especially their thermodynamic topology, has recently received much attention. a comprehensive review of these concepts is in the literature \cite{3000,3001,3002,3003,3004,3005,3006,3007,3008,3009,3010,3011,3012,3013,3014,3015,3016,3017,3018,3019,3020,3021}.\\\\
All the results showed that the inversion curves for all these black hole systems have only positive slopes. We are interested in exploring whether this feature is universal for all black hole systems, or whether other effects can alter it. To this end, we will focus on two types of black holes: AEMPYM and AKS. These black holes have additional parameters that can affect their thermodynamic behavior and phase transitions. AEMPYM are black holes that have a non-linear electromagnetic field, which is described by a power-law function of the field strength\cite{38}. This field can be seen as a generalization of the Maxwell field, which is the standard model of electromagnetism. AKS black holes are black holes that have electric charge and angular momentum in a low-energy limit of heterotic string theory\cite{39}. This theory is a type of string theory that combines the features of bosonic and supersymmetric strings. These black holes have different properties and characteristics than the standard AdS black holes, such as the existence of a dilaton field, which is a scalar field that couples to the electromagnetic field and the curvature. We will investigate how these parameters affect the inversion curves and the JTE process for these black holes, and compare them with the previous results for other black hole systems.\\\\
The structure of this paper is as follows. In Sec. 2, we give a brief overview of the JTE. In Sec. 3 and Sec. 4, we introduce briefly AEMPYM and the AKS black holes and their thermodynamic properties. In Sec. 5, we study the JTE process for these black holes, derive an explicit expression for the JTC, and analyze and discuss the effect of the parameters of each model on the inversion curves. In Sec. 6, we present our conclusion and discussion.
\section{Joule-Thomson expansion}
In classical thermodynamics, the JTE process, discovered in 1852, is a method of cooling or heating a system by changing its pressure and volume, without adding or removing heat. In this process, the high-pressure gas passes through a porous plug into a region with a low pressure, while keeping the enthalpy constant. Since it is a constant-enthalpy process, it can be used to experimentally measure the lines of constant enthalpy on the $(P, T)$ figure of a system. Combined with the specific heat capacity at constant pressure, it allows the complete measurement of the thermodynamic potential for the gas\cite{20.1,28,29,36,36.1,48}.\\\\In this method, The main goal is to investigate the behavior of the coefficient that describes the temperature change during the expansion or compression of a system at constant enthalpy, which is denoted by $\mu$ and is known as the JTC\cite{20.1,28,29,36,36.1,48},
\begin{equation*}\label{eq1}
\mu=\big(\frac{\partial T}{\partial P}\big)_H=\frac{1}{C_{P}}\big[T(\frac{\partial V}{\partial T}\big)_P-V\big],
\end{equation*}
where T, P, V, $C_P$, H are determined by temperature, pressure, volume, specific heat capacity at constant pressure, and enthalpy (the mass of the black hole could be considered as the enthalpy, $M \equiv H$, rather than the internal energy of the gravitational system), respectively. If the above coefficient is positive, as a result of pressure reduction, the temperature will decrease. In other words, the expansion of the gas causes cooling and the compression of the gas under investigation causes it to heat up. In other words, the positive JTC indicates the same direction of temperature and pressure. Whereas, if the JTC is negative, a decrease in pressure causes an increase in temperature.\cite{20.1,28,29,36,48}\\The JT inversion temperature, which is determined by setting  $\mu = 0$, is the temperature at which the sign of the JTC changes. Most real gases have an inversion point. The temperature of this point depends on the gas pressure before expansion\cite{20.1,28,29,36,48}.
\begin{equation*}\label{eq1}
T_{i}=V\big(\frac{\partial T}{\partial V}\big)_P,
\end{equation*}
where $T_i$ is inversion temperature. The important point is that if you plot the JTC on the $(P, T)$ diagram, a closed parabolic curve is created. In simpler terms, the inversion temperature is placed on the boundary of the curve of temperature changes in terms of pressure\cite{20.1,28,29,36,48}. At this temperature, the JTC changes from negative to positive. At a given pressure, the line intersects the drawn curve at two different points. These two points are called low temperature and high temperature of inversion. In fact, at temperatures higher and lower than these two temperatures, the sign of the JTC is negative and between these two temperatures, the sign of the JTC is positive\cite{20.1,28,29,36,48}.
According to the above, the interesting phenomenon in this process is that the (P, T) diagram has two regions: one where the gas cools down and one where the gas heats up\cite{20.1,28,29,36,48}. These regions are separated by the inversion curve, the curve that shows the points where the system temperature does not change during the expansion process, which divides the graph into two regions: the cooling region and the heating region. The cooling region is where the gas temperature decreases as the pressure decreases, and the heating region is where the gas temperature increases as the pressure decreases\cite{20.1,28,29,36,48}.
The inversion curve depends on the type of the gas and its initial conditions. If with respect to zero coefficient for ideal gases, we choose the van der Waals system, which is a more realistic model than the ideal gas, and takes into account the finite size and the attractive forces of the molecules, we find that for the van der Waals system, the inversion curves have both positive and negative slopes, forming a circle in the pressure axis. The inversion curve for the van der Waals system has a negative slope in the low-pressure region, where the attractive forces dominate, and a positive slope in the high-pressure region, where the repulsive forces dominate\cite{20.1,28,29,36,48}.
\section{Case I: AdS-Einstein-Maxwell-Power-Yang-Mills black holes}
AEMPYM black holes are studied within supergravity theories, particularly in AdS space, due to their relevance to string theory and the AdS/CFT correspondence. These black holes are solutions to supergravity equations with additional matter fields like Maxwell and power-Yang-Mills fields, extending the Einstein-Maxwell theory with a negative cosmological constant and a non-Abelian gauge field. The history of AEMPYM black holes dates back to the first black holes in Einstein-Yang-Mills theory by Yasskin and Kasuya. Initially, non-Abelian black holes were unstable in asymptotically flat geometries, but later found stability in AdS space. The Wu-Yang ansatz is a key form for deriving magnetic-type solutions. Recent research focuses on their thermodynamic properties, phase transitions, and connections to gauge/gravity duality. These black holes model strongly coupled systems in dual field theories, offering insights into nonperturbative phenomena. Their rich phase structures and holographic interpretations link gravitational physics with field theory, addressing fundamental questions in quantum gravity and strongly coupled systems\cite{37.1,37.2,37.3,37.4,37.5,37.6,37.9,37.10,37.11,37.12,37.7,37.8,37.13,37.14,37.15,37.16,6000,6001,6002,6003,37.17,37.18,38}.
\\\\ Here, we investigate the thermodynamic properties of N-dimensional gravity theories that combine Einstein's gravity, Maxwell's electromagnetism, and power-Yang-Mills fields. The presence of a cosmological constant $\Lambda$ further enriches the dynamics of these gravitational systems. The theory is grounded in the following action, which encapsulates the fundamental interactions\cite{1},
\begin{equation}\label{1}
\begin{split}
I=\frac{1}{2}\int dx^N \sqrt{-g}\bigg(R-\frac{(N-1)(N-2)\Lambda}{3}-F_{\mu\nu}F^{\mu\nu}-\big(Tr(F_{\mu\nu}^{(a)}F^{(a)\mu\nu})\big)^q\bigg).
\end{split}
\end{equation}
In the context of N-dimensional gravity theories, we encounter the trace element and field definitions that play essential roles in understanding spacetime dynamics,
$Tr(.) = \sum_{a=1}^{(N-1)(N-2)/2}(.)$ . Here, $R$ represents the Ricci scalar, and $q$ is a real positive parameter. The summation involves a combination of indices, capturing the intricate geometric properties of the manifold. Yang-Mills and Maxwell Fields are fundamental in describing the interactions within the gravitational system. The Yang-Mills field corresponds to non-abelian gauge theories, capturing the strong force interactions. The Maxwell field represents electromagnetism, describing the behavior of charged particles. In summary, the trace element and the interplay between Yang-Mills and Maxwell fields contribute to the rich tapestry of N-dimensional gravity. These concepts are crucial for unraveling the underlying physics and exploring the intricate fabric of spacetime. The Yang-Mills and Maxwell fields are defined as \cite{1},
\begin{equation}\label{2}
\begin{split}
F_{\mu\nu}^{(a)}=\partial_{\mu}A_{\nu}^{(a)}-\partial_{\nu}A_{\mu}^{(a)}+\frac{1}{2\sigma}C^{(a)}_{(b)(c)}A^b_{\mu}A^c_{\nu},
\end{split}
\end{equation}
\begin{equation}\label{3}
\begin{split}
F_{\mu\nu}=\partial_{\mu}A_{\nu}-\partial_{\nu}A_{\mu},
\end{split}
\end{equation}
The structure constants of the Lie group $( G )$, which has $((N-1)(N-1)/2)$ parameters, are denoted by $( C^{(a)}_{(b)(c)} )$. The coupling constant is represented by $( \sigma )$, and the Yang-Mills potentials for the $( SO(N-1) )$ gauge group are given by $( A^{(a)}_{\mu} )$. In this context, $( A_{\mu} )$ signifies the conventional Maxwell potential as discussed in reference\cite{38}. The metric solution that corresponds to the line element of an ( N )-dimensional spherically symmetric space is detailed in reference\cite{39}.
\begin{equation}\label{4}
\begin{split}
ds^2=-f(r) dt^2+\frac{1}{f(r)}dr^2+r^2 d\Omega_n^2.
\end{split}
\end{equation}
The term $( d\Omega_n^2 )$ signifies the volume element on an ( n)-dimensional unit sphere. The scope of this study is the EMPYM theory in dimensions $( N )$, where $( N = n + 2 )$ and is at least 4, with the constraint that the parameter $( q )$ is distinct from ( $\frac{n+1}{4} $). q is real positive parameter of AEMPYM black hole. A detailed exposition of the ( N )-dimensional EMPYM black hole solution, which includes a negative cosmological constant and conforms to the condition ( $q \neq \frac{n+1}{4}$ ), is presented in the referenced document\cite{38}.
\begin{equation}\label{5}
\begin{split}
&f(r)=1-\frac{2m}{r^{n-1}}-\frac{\Lambda r^2}{3}+\frac{2(n-1)C^2}{nr^{2(n-1)}}+\frac{Q}{r^{4q-2}}\\
&Q=\frac{[n(n-1)Q_1^2]^q}{n(4q-n-1)}.
\end{split}
\end{equation}
It should be highlighted that within this framework, the variable $( m )$ corresponds to the black hole's mass. The symbols $( C )$ and $( Q_1 )$ are indicative of the charges associated with the Maxwell field and the Yang-Mills field, respectively. When considering the cosmological constant as a thermodynamic variable in an extended phase space, it is interpreted as the thermodynamic pressure $( P )$, defined by the relation $( P = -\frac{\Lambda}{8\pi} )$. Under these stipulations, the expressions for the Hawking temperature, the black hole's mass, and its entropy can be derived as follows\cite{38},
\begin{equation}\label{6}
\begin{split}
T=-\frac{C^2 (n-1)^2}{2 \pi  n r_+^{2 n-1}}+\frac{2}{3} (n+1) P r_+-\frac{Q (-n+4 q-1)}{4 \pi  r_+^{4 q-1}}+\frac{n-1}{4 \pi  r_+},
\end{split}
\end{equation}
\begin{equation}\label{7}
\begin{split}
M=\frac{n \omega_n}{48 \pi } \bigg(\frac{6 C^2 (n-1)}{n r_+^{n-1}}+8 \pi  P r_+^{n+1}+3 Q r_+^{n-4 q+1}+3 r_+^{n-1}\bigg),
\end{split}
\end{equation}
\begin{equation}\label{8}
\begin{split}
S=\frac{\omega_n r_+^n}{4}, \qquad    \omega_n=\frac{2 \pi ^{\frac{n+1}{2}}}{\Gamma \left(\frac{n+1}{2}\right)},
\end{split}
\end{equation}
where $S$ is entropy. In this context, the term $( r_+ )$ signifies the radius of the black hole's event horizon, while $( \omega_n )$ represents the volume of an ( n )-dimensional unit sphere. To determine the critical points, we can utilize the relations $\frac{\partial T}{\partial r_{+}}=\frac{\partial^2 T}{\partial r_{+}^2}=0$. Consequently, the critical temperature and pressure for the AEMPYM black hole are calculated as follows\cite{38},
$$T_c=\frac{r_+^{-2 n-4 q-1}\bigg((n-1) r^{4 q} \big(r_+^{2 n}-2 C^2 (n-1) r_+^2\big)-2 q Q (-n+4 q-1) r_+^{2 n+2}\bigg)}{2 \pi }$$
and
$$P_c=\frac{3 \bigg(\frac{(n-1) r_+^{-2 (n+1)} \big(n r_+^{2 n}-2 C^2 (2 n^2-3 n+1) r^2\big)}{n}+(4 q-1) Q (n-4 q+1) r_+^{-4 q}\bigg)}{8 \pi  (n+1)}$$
\section{Case II: AdS-Kerr-Sen black hole}
The KS black hole, a rotating and charged black hole from heterotic string theory, generalizes the Kerr-Newman-AdS black hole. It includes a dilaton and an axion field, which are scalar and pseudoscalar fields from string theory. The Schwarzschild solution, discovered in 1916, describes a simple black hole, while the Kerr solution in 1963 introduced rotating black holes. The Kerr-Newman black hole, found in 1965, added charge, and Ashoke Sen extended it to include dilaton and axion fields, resulting in the Kerr-Sen black hole. The AKS black hole retains some properties of the Kerr-Newman-AdS black hole but has unique features like mass and angular momentum dependence on the dilaton charge. It also sometimes violates the cosmic censorship conjecture. The KS black hole's properties and thermodynamics have been extensively studied, especially in the context of the AdS/CFT correspondence. Research has focused on their thermodynamic properties, stability, and connections to dual-field theories. The holographic interpretation of KS black holes has linked gravitational physics with field theory, addressing fundamental questions in quantum gravity and strongly coupled systems. This ongoing research continues to explore the dynamic behavior, evolution, and implications for chaos and information loss puzzles in quantum gravity\cite{40,1200,1201,40.1,40.2,40.3,40.4,40.5,40.6,40.7,40.8,40.9,40.10,40.11}.\\ Sen's work \cite{40} unveiled a charged, rotating black hole solution within the framework of low-energy heterotic string theory, termed the Kerr-Sen black hole. This solution modifies the conventional action of general relativity by incorporating supplementary fields derived from heterotic string theory, as delineated by\cite{40.1,40.6},
\begin{equation}\label{9}
\begin{split}
S = \int d^4 x \sqrt{-\widetilde{g}}e^{-\Phi} \left[\mathcal{R} + (\nabla \Phi)^2 - \frac{1}{8} F^2 - \frac{1}{12}H^2 \right],
\end{split}
\end{equation}
In the given context, $\tilde{g}$ represents the determinant of the metric tensor, denoted as $g_{\mu\nu}$. The Ricci scalar is symbolized by $\mathcal{R}$, which is a key scalar quantity in General Relativity that represents curvature. The term $F$ is defined as $F_{\mu\nu}F^{\mu\nu}$, where $F_{\mu\nu}$ is the field strength tensor associated with the $U(1)$ Maxwell field, encapsulating the electromagnetic field's properties. The scalar dilaton field is represented by $\Phi$, which arises in theories of gravity and string theory as a means of introducing a varying gravitational constant. Lastly, $H$ is expressed as $H_{\mu\nu\rho}H^{\mu\nu\rho}$, indicating the field strength of the axion field, a hypothetical elementary particle postulated to account for the CP problem in quantum chromodynamics. A conformal transformation of the metric is employed to transition to the Einstein frame, which is a particular representation of the gravitational field equations where the tensorial part of the action takes the simplest form. This transformation adjusts the metric tensor while preserving the angle between vectors, thereby simplifying the action's structure in relation to the curvature of spacetime\cite{40.1,40.6},
\begin{equation}\label{10}
\begin{split}
ds^{2}_{E}=e^{-\Phi}d\widetilde{s}^2.
\end{split}
\end{equation}
So, we will have $S$ in the Einstein frame as follows\cite{40.1,40.6},
\begin{equation}\label{11}
\begin{split}
S = \int d^4 x \sqrt{-g} \left[R - \frac{1}{2}(\nabla \Phi)^2 -\frac{e^{-\Phi}}{8} F^2 - \frac{e^{-2\Phi}}{12} H^2 \right].
\end{split}
\end{equation}
In the absence of dilaton, vector, and axion fields, the Einstein-Hilbert action is a specific case. To derive black holes with a nonzero cosmological constant, the three-form field $( H \equiv d\mathcal{B} - A \wedge \frac{F}{4} = -e^{2\phi} \star d\chi )$ can be reformulated, where $\mathcal{B}$ is an anti-symmetric two-form potential and the $( \star )$ operator represents the Hodge duality. As a result, the Lagrangian can be written in an alternative but equivalent form\cite{900,40.1},
$$
\hat{L} = \sqrt{-g} \left[ R - \frac{1}{2} (\partial \phi)^2 - \frac{1}{2} e^{2\phi} (\partial \chi)^2 - e^{-\phi} F^2 \right] + \frac{\chi}{2} \epsilon^{\mu\nu\rho\lambda} F_{\mu\nu} F_{\rho\lambda} + \sqrt{-g} \left[ 4 + e^{-\phi} + e^{\phi} (1 + \chi^2) \right] / \ell^2,
$$
where $( \epsilon^{\mu\nu\rho\lambda} )$ is the four-dimensional Levi-Civita antisymmetric tensor density, and $( \ell )$ denotes the cosmological scale or the reciprocal of the gauge coupling constant.
In Boyer-Lindquist coordinates, the metric for an AKS black hole is described by the following expression\cite{40.1,40.6},
\begin{equation}\label{12}
\begin{split}
ds^2 = -\frac{\Delta_r}{\rho^2}(dt - \frac{a \sin^2 \theta}{\Xi} d\phi)^2 + \frac{\rho^2}{\Delta_r} dr^2 + \frac{\rho^2}{\Delta_\theta} d\theta^2 + \frac{\sin^2 \theta\Delta_\theta}{\rho^2}[-\frac{(r^2 +2br+ a^2)}{\Xi}d\phi + a dt]^2,
\end{split}
\end{equation}
where
\begin{equation}\label{13}
\begin{split}
&\rho^2= r^2 + a^2 \cos^2\theta +2br,\\
&\Delta= (r^2 +2br+ a^2)(1 + \frac{r^2+2br }{\ell^{2}})- 2mr,\\
&\Xi=1- \frac{a^2}{\ell^{2}},\\
&\Delta_{\theta}=1- \frac{a^2}{\ell^{2}}\cos^{2}\theta.
\end{split}
\end{equation}
The $b$ parameter signifies the dyonic charge of black holes ($b = \widehat{q}^2/(2m)$), linked to their electric charge ($\widehat{q}$) and mass (m). $a$ is the rotation parameter. The Kerr-Sen black holes emerge in certain limits, while the GMGHS solution describes non-rotating cases. Pioneering work by Gibbons, Maeda, Garfinkle, Horowitz, and Strominger has led to a deeper understanding of these black holes, especially in the context of AdS spacetimes, where their mass, angular momentum, and charge are interrelated. Studies have focused on their shadows and thermodynamics, contributing to the broader field of black hole physics\cite{41,42,43,44,45}. The $M$, $J$, and $Q$ are given by,
\begin{equation}\label{14}
\begin{split}
&M = \frac{m}{\Xi^2},\\
&J = \frac{ma}{\Xi^2},\\
&Q = \frac{\widehat{q}}{\Xi}.
\end{split}
\end{equation}
Also, we can obtain the entropy, mass, and temperature in terms of $r_+$ and $a$, $b$ as follows\cite{40.1},
\begin{equation}\label{15}
\begin{split}
S = \frac{A}{4} = \frac{\pi (r_+^2+2br + a^2)}{\Xi},
\end{split}
\end{equation}

\begin{equation*}\label{15555}
\begin{split}
M=\frac{\left(a^2+2 b r+r^2\right) \left(\frac{2 b r+r^2}{l^2}+1\right)}{2 \Xi^2 r},
\end{split}
\end{equation*}
\begin{equation*}\label{15555}
\begin{split}
T=\frac{a^2 \left(r^2-l^2\right)+r^2 \left(4 b^2+8 b r+l^2+3 r^2\right)}{4 \pi  l^2 r \left(a^2+2 b r+r\right)}
\end{split}
\end{equation*}
In the context of black hole physics, the variables $ r_+ $, $M$, $J$, and $Q$ represent the event horizon radius, mass, angular momentum, and charge, respectively. The event horizon radius $ r_+ $ is specifically the largest solution to the equation $ \Delta = 0 $. Setting the parameter b to zero aligns the thermodynamic properties with those of the Kerr-AdS black holes, which are characterized as rotating solutions to the four-dimensional Einstein field equations with a negative cosmological constant. Furthermore, the critical temperature and pressure for the AKS black hole have been determined through specific calculations as follows\cite{45},
\begin{equation*}\label{15}
\begin{split}
&T_c=-192 J^2 Q^2 \bigg(\pi  (26 v^2+14 v-33)-27\bigg) v^3\\
&+\bigg(v^4 \bigg[36864 J^4 \bigg(153 Q^4 v^2+\pi ^2 Q^4 v^2 \big(26 v^2+14 v-33\big)^2-6 \pi  Q^4 v^2 \big(90 v^2+206 v-425\big)+72 \pi  Q^2\bigg)\\&+384 \pi  J^2 Q^2 v^4 \bigg(\pi  \big(12 Q^2-1\big) \big(26 v^2+14 v-33\big)-228 Q^2+3\bigg)+\pi ^2 (1-12 Q^2)^2 v^6\bigg]\bigg)^{1/2}\\
&+12 \pi  Q^2 v^5-\pi  v^5\bigg/\bigg(4608 \pi  J^2 Q^2 v^4\bigg),
\end{split}
\end{equation*}
and
\begin{equation*}\label{15}
\begin{split}
&P_c=\frac{(9+11 \pi ) \bigg[J^2 Q^2(12 Q^2-1) \bigg(-4 (19+33 \pi ) Q^2+11 \pi +1\bigg) v^9 \bigg(768 J^2 Q^2 (13 v+7)+(12 Q^2-1) v\bigg)\bigg]^{1/2}}{\sqrt{2} \pi  v^7}\\&+\bigg[12 (416 J^2+1) Q^2-1\bigg] \bigg(v^8 \pi  (12 Q^2-1) v \bigg[768 J^2 Q^2 (13 v+7)+12 Q^2 v-1\bigg]\\&+1152 J^2 \bigg(-4 (19+33 \pi ) Q^2+11 \pi +1\bigg) Q^2\bigg)^{1/2}\bigg/\bigg(9216 \sqrt{\pi } J^2 Q^2 v^5\bigg)\\&+\frac{(12 Q^2-1) \big(12 (832 J^2+1) Q^2-1\big)}{9216 J^2 Q^2}-\frac{(9+11 \pi ) (12 Q^2-1)}{24 \pi  v^2}+\frac{7 (12 Q^2-1)}{24 v}.
\end{split}
\end{equation*}
where $v$ is the new variable which relates to the specific volume of the van der Waals fluid as\cite{45},
\begin{equation*}\label{15}
\begin{split}
v=2\big(\frac{3V}{4\pi}\big)^{1/3}=2r_++2b+\frac{12J^2}{\big(8\pi Pr_+^4+3r_+^2\big)^2}\big(4b+\frac{80}{3}Pbr_+^2+8\pi Pr_+^3+3r_+\big).
\end{split}
\end{equation*}
\section{Discussion and results}
In this section, we investigate the JTE process for AdS black holes: AEMPYM and AKS and derive a clear formula for the JTC. We also examine how the charge and the parameters of AEMPYM and Kerr-Sen theories affect the inversion curves. We also compare the inversion curves for different scenarios
\subsection{Case I}
\begin{figure}[H]
 \begin{center}
 \subfigure[]{
 \includegraphics[height=4.5cm,width=4.5cm]{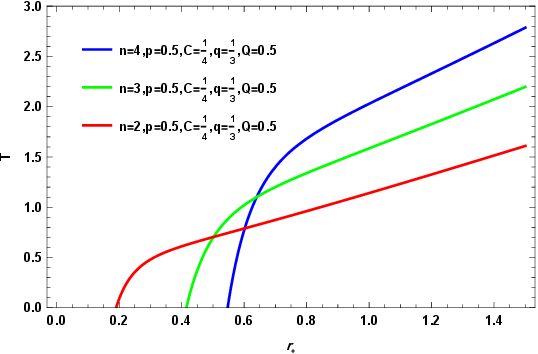}
 \label{1a}}
 \subfigure[]{
 \includegraphics[height=4.5cm,width=4.5cm]{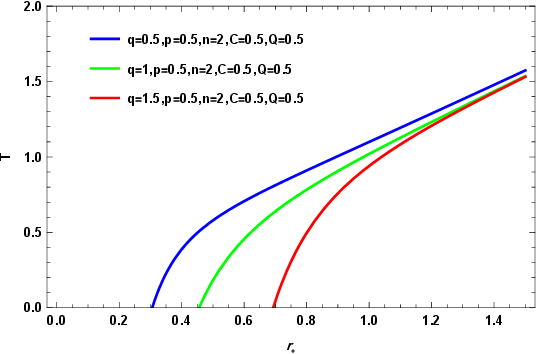}
 \label{1b}}
 \subfigure[]{
 \includegraphics[height=4.5cm,width=4.5cm]{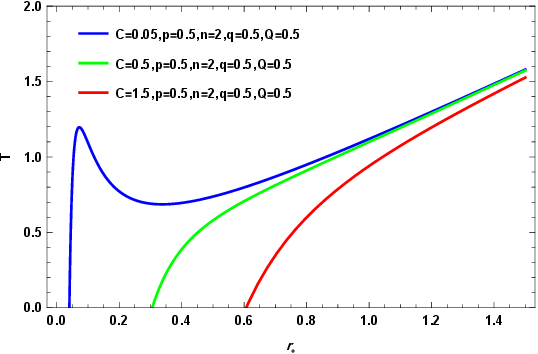}
 \label{1c}}
 \subfigure[]{
 \includegraphics[height=4.5cm,width=4.5cm]{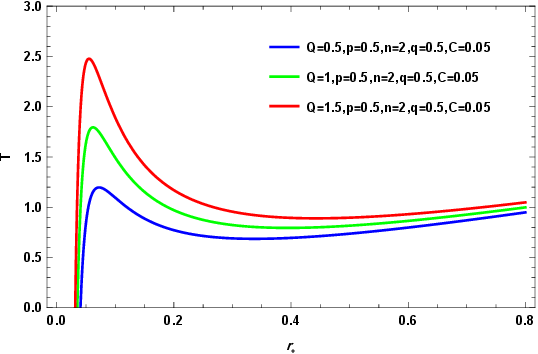}
 \label{1d}}
  \caption{\small{The plot of temperature $T$ in terms of $r_+$ at $P =0.5, C=1/4, q=1/3$, and $Q=0.5$ for different values of $n$ in Fig. (\ref{1a}), at $P =0.5, n=2, C=0.5$, and $Q=0.5$ for different values of $q$ in Fig. (\ref{1b}), at $P =0.5, n=2, q=0.5$, and $Q=0.5$ for different values of $C$ in Fig. (\ref{1c}) and at $P =0.5, n=2, q=0.5$, and $C=0.05$ for different values of $Q$ in Fig. (\ref{1d}) for the AEMPYM black hole}}
 \label{m1}
 \end{center}
 \end{figure}

\begin{figure}[H]
 \begin{center}
 \subfigure[]{
 \includegraphics[height=7.5cm,width=8cm]{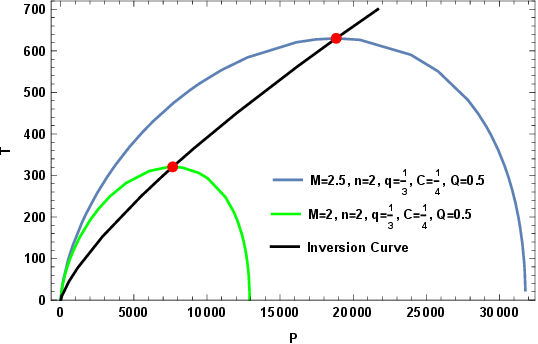}
 \label{2a}}
 \subfigure[]{
 \includegraphics[height=7.5cm,width=8cm]{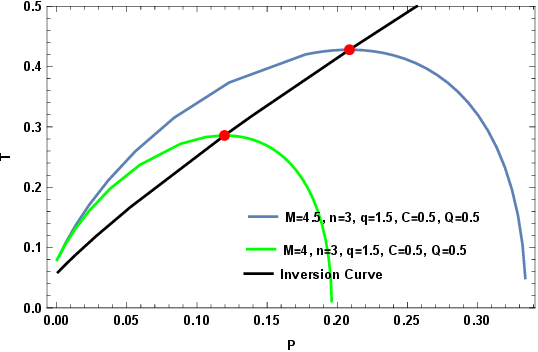}
 \label{2b}}
  \caption{\small{The plot of isenthalpic and inversion curves for different values of $M$. In Fig. (\ref{2a}) at $n =2, C=1/4, Q=0.5$, and $q=1/3$, in Fig. (\ref{2b}) at $n=3, q =1.5, C=0.5$, and $Q=0.5$ for the AEMPYM black hole}}
 \label{m2}
 \end{center}
 \end{figure}

\begin{figure}[H]
 \begin{center}
 \subfigure[]{
 \includegraphics[height=4.5cm,width=4.5cm]{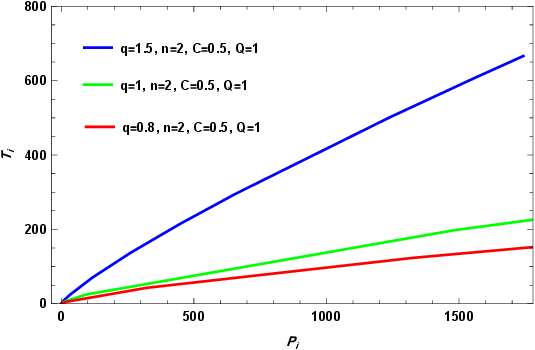}
 \label{3a}}
 \subfigure[]{
 \includegraphics[height=4.5cm,width=4.5cm]{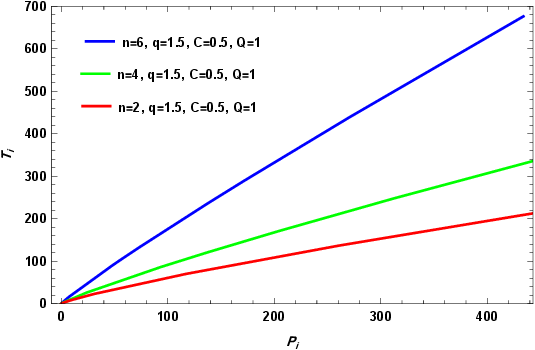}
 \label{3b}}
 \subfigure[]{
 \includegraphics[height=4.5cm,width=4.5cm]{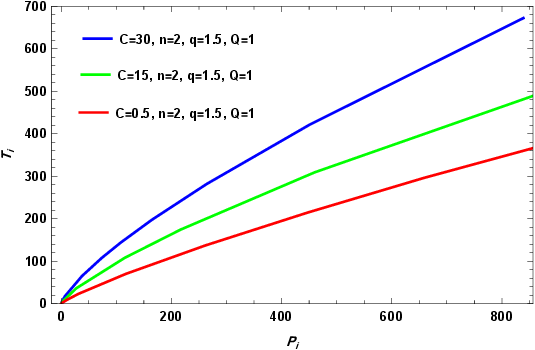}
 \label{3c}}
 \subfigure[]{
 \includegraphics[height=4.5cm,width=4.5cm]{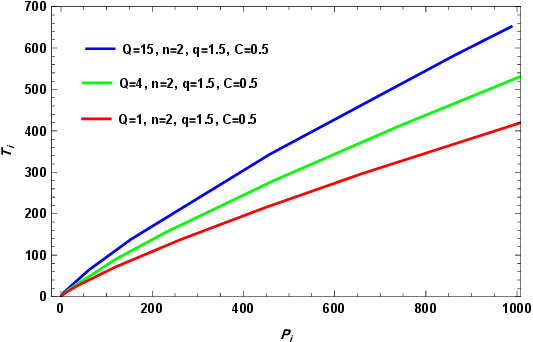}
 \label{3d}}
  \caption{\small{The plot of inversion curves ($T_i-P_i$). In Fig. (\ref{3a}) at $n =2, C=0.5$, and $Q=1$ for different values of $q$, in Fig. (\ref{3b}) at $q =1.5, C=0.5$, and $Q=1$ for different values of $n$, in Fig. (\ref{3c}) at $n=2, q=1.5$, and $Q=1$ for different values of $C$  and in Fig. (\ref{3d}) at $n=2, q=1.5$, and $C=0.5$ for different values of $Q$  for the AEMPYM black hole}}
 \label{m3}
 \end{center}
 \end{figure}

\begin{figure}[H]
 \begin{center}
 \subfigure[]{
 \includegraphics[height=5cm,width=8.5cm]{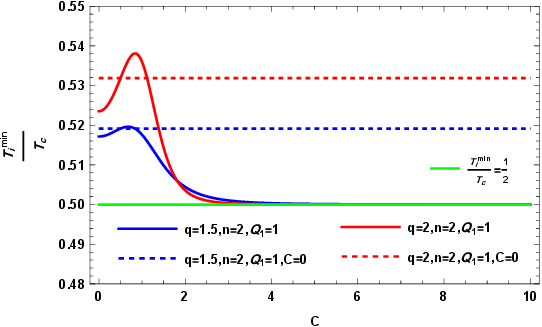}
 \label{4a}}
  \caption{\small{The plot of $(\frac{T_i^{min}}{T_c})$ in terms of $C$. The solid blue line illustrated with $q=1.5, n=2$, and $Q_1=1$, solid red line denoted with $q=2, n=2$, and $Q_1=1$, dashed blue line represented with $q=1.5, n=2,  Q_1=1$ and $C=0$, and dashed red line corresponds with $q=2, n=2, Q_1=1$ and $C=0$ for the AEMPYM black hole }}
 \label{m4}
 \end{center}
 \end{figure}

In this section, we explore the JTE of black hole systems in the extended phase space, where the black hole mass M is the same as the enthalpy H. In our study, we conduct a short comparison of our results with the (JT) process as applied to other black holes. These comparative results are summarized in table 5. This allows us to evaluate the consistency and discrepancies between our findings and the other data within this concept. Using \cite{36,37.4} and considering the mass of black holes, we can write the pressure P as a function of M and $r_+$, the mass and the horizon radius of a black hole. By substituting this expression in the temperature formula, we can obtain the temperature as a function of M and $r_+$ as well. Furthermore, we can express the mass M and the temperature T in terms of the pressure P and the radius $r_+$ of a black hole. Therefore, we get,
\begin{equation}\label{16}
P(M,r_+)=-\frac{3 \left(\frac{2 C^2 (n-1) r_+^{-2 n}}{n}+Q r_+^{-4 q}+\frac{1}{r_+^2}\right)}{8 \pi }+\frac{6 M r_+^{-n-1}}{n \omega_n }.
\end{equation}
Also, we have $T$ in terms of $M$ and $r_+$ as follows,
\begin{equation}\label{17}
T(M,r_+)=-\frac{2 C^2 (n-1) r_+^{2-2 n}+2 q Q r_+^{2-4 q}+1}{2 \pi  r_+}+\frac{4 M (n+1) r_+^{-n}}{n \omega_n }.
\end{equation}
Due to the complexity of the aforementioned such as $T(M, P)$ equations, an analytical solution is not feasible. Therefore, we have resorted to numerical solutions to further our work. A portion of these calculations can be found in the appendix. According to the definition of the JTC $\mu = \left( \frac{\partial T}{\partial P} \right)_M$, the inversion pressure and temperature between the cooling and heating regions are ($\tilde{P}$-$\tilde{T}$), which are determined by $\mu = 0$. Therefore, the most important thing is to find the function expression of $\mu$. By setting $\mu = 0$, one can obtain the inversion points ($\tilde{P}$, $\tilde{T}$) for different fixed enthalpy M. With respect to\cite{21}, the JTC is given by,
\begin{equation}\label{18}
\begin{split}
\mu=\big(\frac{\partial T}{\partial P}\big)_M=\frac{1}{C_{P}}\big[T(\frac{\partial V}{\partial T}\big)_P-V\big].
\end{split}
\end{equation}
This approach is elegant. However, in this paper, we will use more straightforward methods by applying only mass and temperature to derive the JTC $\mu$. we can see that temperature is a function of pressure and radius, and radius is a function of pressure and mass. So, the JTC is given by,
\begin{equation}\label{19}
\begin{split}
\mu=\big(\frac{\partial T}{\partial P}\big)_M=\frac{\partial_{,r_+} T}{\partial_{,r_+} P}.
\end{split}
\end{equation}
Now, using the relationship $\mu=\frac{\partial T}{\partial P}|_M=\frac{\frac{\partial T}{\partial r_+}}{\frac{\partial P}{\partial r_+}}$, we calculate Joule Thomson coefficient as follows,
\begin{equation}\label{20}
\begin{split}
&\mu=\frac{2 n r_+\bigg(\omega _n \bigg[r_+^{4 q} \big(2 C^2 (2 n^2-3 n+1) r_+^2+r_+^{2 n}\big)+8 q^2 Q r^{2 n+2}-2 q Q r_+^{2 n+2}\bigg]-8 \pi  M (n+1) r_+^{n+4 q+1}\bigg)}{3 \bigg(n \omega _n \bigg[r_+^{4 q} \big(2 C^2 (n-1) r_+^2+r_+^{2 n}\big)+2 q Q r_+^{2 n+2}\bigg]-8 \pi  M (n+1) r^{n+4 q+1}\bigg)}.
\end{split}
\end{equation}
When the value of the coefficient $\mu$ is positive during the expansion, it means that the temperature decreases and therefore it is called a cooling phenomenon. However, when $\mu$ is negative, the temperature increases, and this is called a heating process. Using various equations, we can write the mass M and the temperature T as functions of the pressure P and the radius $r_+$, which are the properties of a black hole.
For $\mu = 0$, we can obtain the inversion temperature, in which the process of the temperature changes reverses. It can be obtained by the formula,
\begin{equation}\label{21}
T_{i}=V\big(\frac{\partial T}{\partial V}\big)_P,
\end{equation}
\begin{equation}\label{22}
\begin{split}
V=\frac{\partial M}{\partial P}=\frac{1}{6} n r_+^{n+1} \omega _n.
\end{split}
\end{equation}
At the inversion temperature, the value of $\mu$ is 0, and the inversion temperature is determined by the following equation:
\begin{equation}\label{23}
\begin{split}
T_i=V \frac{\partial T}{\partial V}=V\frac{ \frac{\partial T}{\partial r_+}}{\frac{\partial V}{\partial r_+}}.
\end{split}
\end{equation}
This is beneficial for identifying the areas of heating and cooling in the $T-P$ plane. We calculate $T_i$ using the Eqs. (\ref{17}), (\ref{20}), (\ref{21}), (\ref{22}) and (\ref{23}),
\begin{equation}\label{24}
\begin{split}
T_i=\frac{r_+ \left(\frac{6 C^2 (n-1)^2 (2 n-1) r_+^{-2 n}}{n}+8 \pi  (n+1) P_i+3 (1-4 q) Q (n-4 q+1) r_+^{-4 q}+\frac{3-3 n}{r_+^2}\right)}{12 \pi  (n+1)}.
\end{split}
\end{equation}
We can also have from Eq. (\ref{17}),
\begin{equation}\label{25}
\begin{split}
T_i=\frac{-\frac{6 C^2 (n-1)^2 r_+^{1-2 n}}{n}+8 \pi  (n+1) r_+ P_i+3 Q (n-4 q+1) r_+^{1-4 q}+\frac{3 (n-1)}{r_+}}{12 \pi }.
\end{split}
\end{equation}
Based on the above equation, we can derive an expression commonly denoted as $( T^{min}_{i} )$. This is achieved by setting the value of $( P_{i} )$ to zero. Also, we can rewrite the Eqs. (\ref{24}) and (\ref{25}) with respect to Eq. (\ref{5}) as follows,
\begin{equation*}\label{25}
\begin{split}
T^{min}_{i} =\frac{r_+^{-2 n-4 q-1} \left(2 C^2 (n-1)^2 (2 n-1) r_+^{4 q+2}+(4 q-1) r_+^{2 n+2} \left((n-1) n Q_1^2\right){}^q-(n-1) n r_+^{2 (n+2 q)}\right)}{4 \pi  n (n+1)},
\end{split}
\end{equation*}
and
\begin{equation*}\label{25}
\begin{split}
T^{min}_{i}=\frac{r_+^{-2 n-4 q-1} \left(-2 C^2 (n-1)^2 r_+^{4 q+2}+r_+^{2 n+2} \left(-\left((n-1) n Q_1^2\right){}^q\right)+(n-1) n r_+^{2 (n+2 q)}\right)}{4 \pi  n}.
\end{split}
\end{equation*}
Fig. (\ref{m1}) displays the Hawking temperature as a function of the horizon. We keep the free parameters constant for each plot. In each subfigure, we can observe some zero points for different free parameters. These zero points correspond to the divergence points of the JTC, as we can easily see in Fig. (\ref{m1}). According to the radius of the horizon and the values of free parameters of a black hole, for small values for the radius of the event horizon, our structural behavior is completely distinct, and for larger radii, the figures almost converge. The differences between our research and the referenced study are discernible. For instance, while our work presents findings in a specific context, the \cite{38} may approach the subject from a different angle. By carefully referencing the other study, we not only acknowledge its contributions but also reinforce the originality and integrity of our research
We have plotted the isenthalpic curves and the inversion curve of the AEMPYM black hole for various values of the free parameter in each plot of Fig. (\ref{m2}). In every subfigure, two isenthalpic curves with different values of $M$, along with the corresponding inversion curve occur at the highest point of the isenthalpic curves. We denote the inversion temperature and pressure of each isenthalpic curve as $T_i$ and $P_i$, respectively. The isenthalpic curves are divided into two regions by the inversion curve: for $P < P_i$, the isenthalpic curve has a positive slope, indicating that the black hole undergoes cooling during the expansion process. However, for $P > P_i$, the isenthalpic curve has a negative slope, implying that the black hole experiences heating during the expansion process. This behavior is consistent for different values of the free parameter in Figs. ((\ref{2a})-(\ref{2b})).
However, we can illustrate the relation between the inversion temperature $T_i$ and the inversion pressure $P_i$ by using the equations given above. Fig. (\ref{m3}) shows the inversion curves for different free parameters.
The inversion curve has only one branch. The inversion temperature rises steadily with the inversion pressure, but the slope of the inversion curves becomes smaller. We can also notice some fine structures in the cases of subfigures (\ref{m3}). For low pressure, the inversion temperature varies with the free parameters that are specified in each plot. It was interesting for us to study the effect of parameters and we observed that the slope of the inversion curve increases with the changes in the various values of each of the parameters.\\
Before the end of this section, if we want to discuss purely the effect of Maxwell charges as a specific result of our work, we should note that in \cite{36.1}, the authors found that for ($ q > 1$ ) and ( $n = 2$ ), the ratio ($ T_i^{min} / T_c $) is fully established. However, in the present work, by adding the Maxwell charge to the mentioned black hole, we found that for ( $C \gg 1$ ), under the condition that we still have had black hole structures (i.e., they did not convert into a singularity system), the Yang-Mills parameter has less effect on the ratio. In any case, ($ T_i^{min} / T_c$ ) is equal to ( 1/2 ), which exhibits the same behavior as a four-dimensional charged black hole.
For $q>1$, when $C$ increases, the $T_{min}/T_{c}$ approaches the value of $1/2$. Also, according to Fig. (\ref{m4}), we can see that for q>1 in the condition that there is no Maxwell charge, this ratio will be a constant value that is greater than 0.5, and this ratio will increase with the increase of q (in the figure as the dashed line is displayed). However, for $q<1$, the conditions are slightly different from the previous situation, because when $q=1/2$, $n=2$, and $Q=1$, the $T_i^{min}/T_{c}$ is equal to $1/2$ for all values of $"C"$, which is the same as a charged black hole in four dimensions.
Also, according to tables (\ref{n1}) to (\ref{n3}), it can be inferred that when $1/2< q< 1$, with the increase of  Maxwell charge, i.e. $C$, the $T_i^{min}/T_{c}$ only approaches the value of $1/2$ and exhibits more similar behavior to a four-dimensional charged black hole.
If for $0<q<1/2$, the $T_i^{min}/T_{c}$ becomes further or smaller than the value of $1/2$, which is contrary to a charged black hole in four dimensions.
In general, when $q>1$, for the increase of Maxwell's charge $C$, the $T_i^{min}/T_{c}$ is almost equal to $1/2$ and when $q=1/2$, for all values of the Maxwell's charge, the $T_i^{min}/T_{c}$ is equal to $1/2$. Also, when $1<q<1/2$, it is close to the value of $1/2$ and finally when $1/2<q<0$, the values of the $T_i^{min}/T_{c}$ become smaller than of $1/2$. The proximity of the ratio $( T_{min}^i /T_c ) $ to 1/2 for different values of ( q ) in the AEMPYM black hole indicates a specific thermodynamic behavior related to the phase transitions and stability of the black hole. When this ratio is close to 1/2, it suggests that the black hole undergoes a Joule-Thomson expansion where the inversion temperature is approximately half of the critical temperature. This behavior can be understood as a balance between the attractive and repulsive forces within the black hole system, influenced by the Maxwell and Yang-Mills fields. The parameter ( q ) affects the strength and nature of these fields, thereby impacting the thermodynamic properties and phase transitions. In the context of the AEMPYM black hole, the Maxwell and Yang-Mills fields play a crucial role in determining the thermodynamic behavior. The parameter (q) essentially modulates the influence of these fields, leading to variations in the inversion temperature. When ( q) is greater than 1, the ratio $( T_{min}^i /T_c )$ approaches 1/2 as Maxwell's charge (C) increases, indicating a stable phase transition. For (q = 1/2), the ratio remains exactly 1/2 for all values of (C), suggesting a consistent thermodynamic behavior. However, for values of ( q ) between 0 and 1/2, the ratio decreases, moving away from 1/2, which implies a shift in the balance of forces within the black hole system, potentially leading to different phase transition characteristics.

\subsection{Case II}
\begin{figure}[H]
 \begin{center}
 \subfigure[]{
 \includegraphics[height=4.5cm,width=4.5cm]{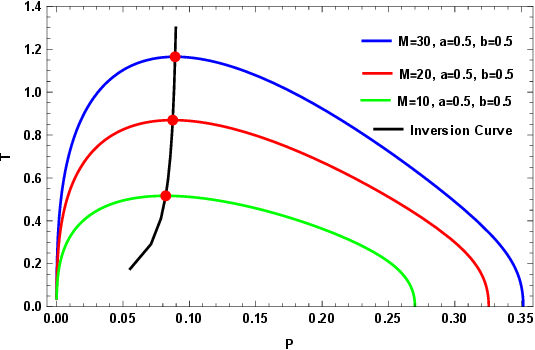}
 \label{5a}}
 \subfigure[]{
 \includegraphics[height=4.5cm,width=4.5cm]{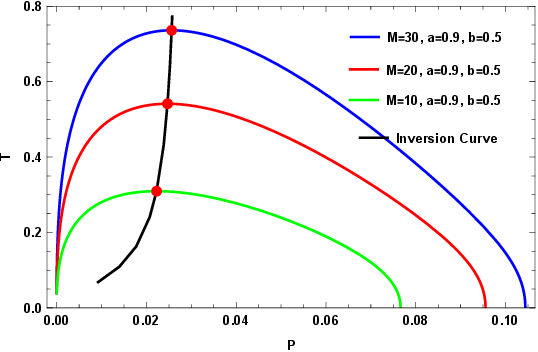}
 \label{5b}}
 \subfigure[]{
 \includegraphics[height=4.5cm,width=4.5cm]{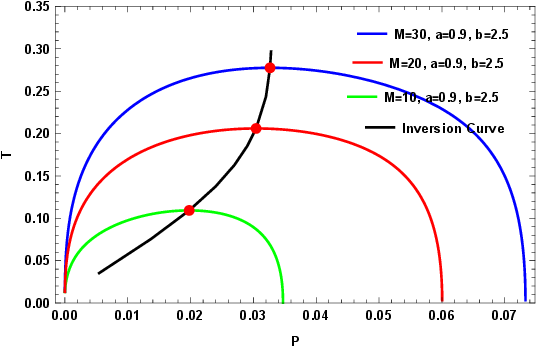}
 \label{5c}}
 \subfigure[]{
 \includegraphics[height=4.5cm,width=4.5cm]{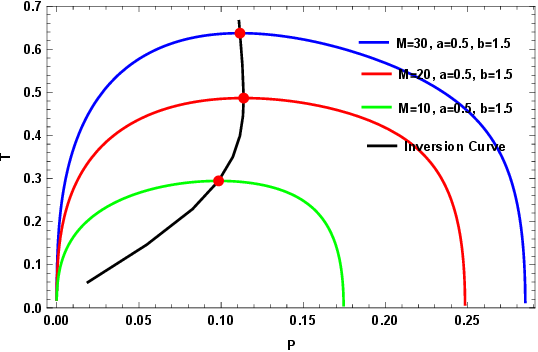}
 \label{5d}}
  \caption{\small{The plot of isenthalpic and inversion curves for different values of $M$. In Fig. (\ref{5a}) at $a=0.5$ and $b=0.5$, in Fig. (\ref{5b}) at $a=0.9$ and $b=0.5$, in Fig. (\ref{5c}) at $a=0.9$ and $b=2.5$, and in Fig. (\ref{5d}) at $a=0.5$ and $b=1.5$ for the AKS black hole}}
 \label{m5}
 \end{center}
 \end{figure}

\begin{figure}[H]
 \begin{center}
 \subfigure[]{
 \includegraphics[height=4.5cm,width=4.5cm]{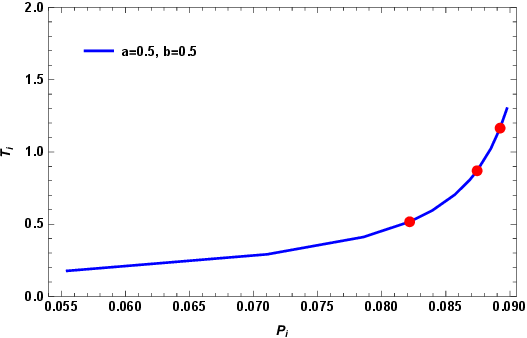}
 \label{6a}}
 \subfigure[]{
 \includegraphics[height=4.5cm,width=4.5cm]{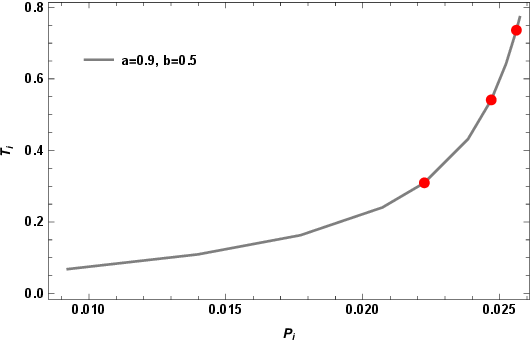}
 \label{6b}}
 \subfigure[]{
 \includegraphics[height=4.5cm,width=4.5cm]{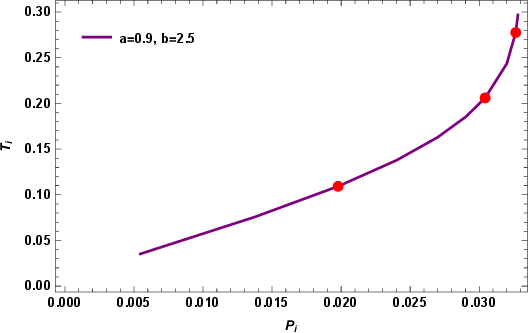}
 \label{6c}}
 \subfigure[]{
 \includegraphics[height=4.5cm,width=4.5cm]{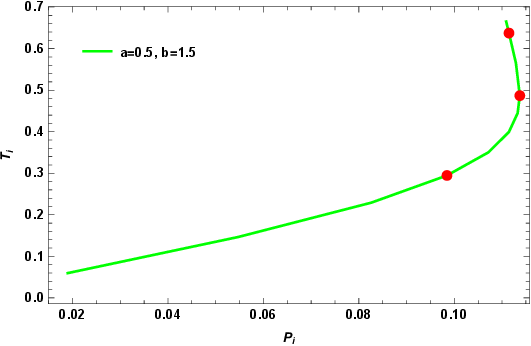}
 \label{6d}}
  \caption{\small{The plot of inversion curves $(T_i-P_i)$ at $a=0.5, b=0.5$ in Fig. (\ref{6a}), at $a=0.9, b=0.5$ in Fig. (\ref{6b}), at $a=0.9, b=2.5$, in Fig. (\ref{6c}), and at $a=0.5, b=1.5$ in Fig. (\ref{6d}) with respect to ($M=10,20,30$)}}
 \label{m6}
 \end{center}
 \end{figure}

\begin{figure}[H]
 \begin{center}
 \subfigure[]{
 \includegraphics[height=4.5cm,width=4.5cm]{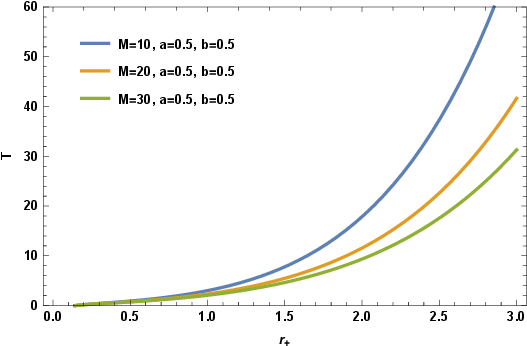}
 \label{7a}}
 \subfigure[]{
 \includegraphics[height=4.5cm,width=4.5cm]{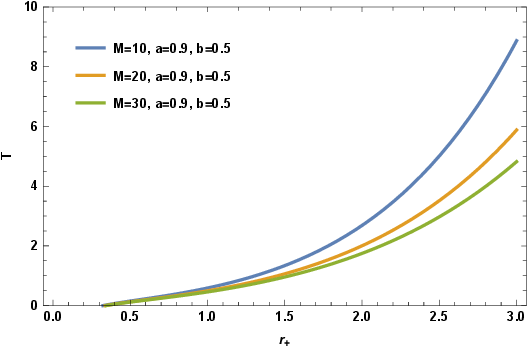}
 \label{7b}}
 \subfigure[]{
 \includegraphics[height=4.5cm,width=4.5cm]{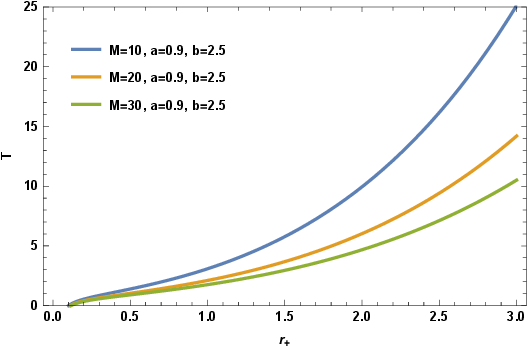}
 \label{7c}}
 \subfigure[]{
 \includegraphics[height=4.5cm,width=4.5cm]{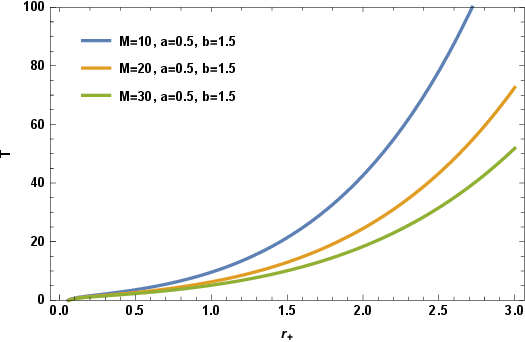}
 \label{7d}}
  \caption{\small{The plot of temperature $T$ in terms of $r_+$ at $a=0.5, b=0.5$ in Fig. (\ref{7a}), at $a=0.9, b=0.5$ in Fig. (\ref{7b}), at $a=0.9, b=2.5$, in Fig. (\ref{7c}), and at $a=0.5, b=1.5$ in Fig. (\ref{7d}) for different values of $M$ for the AKS black hole}}
 \label{m7}
 \end{center}
 \end{figure}

With respect to \cite{40.1,40.6} and the Mass of black holes, We can express pressure P in terms of M and $r_+$ and replace it in the formula for the temperature, which will also become in terms of M and $r_+$. We also rewrite the mass M and temperature T as a function of the pressure P and the radius $r_+$ of a black hole. So we will have,
\begin{equation}\label{25}
P(M,r_+)=-\frac{3\big(a^2+r_+ (2 b-2 \Xi^2 M+r_+)\big)}{8 \pi  r_+ (2 b+r_+) \big(a^2+r_+ (2 b+r_+)\big)}.
\end{equation}
Also, the temperature in terms of M and $r_+$ is as follows\cite{40.1,40.6},
\begin{equation}\label{26}
\begin{split}
&\mathcal{X}=a^4 (b+r_+)+a^2 r_+\big(4 b^2+6 b r_++r_+ (2 r_+-\Xi^2 M)\big)+r_+^2 (2 b+r_+) \big[3 r_+ (b-\Xi^2 M)+2 b (b-\Xi^2 M)+r_+^2\big]\\
&\mathcal{Y}=2 \pi  r_+ (2 b+r_+) (a^2+2 b r_++r_+) (a^2+r_+ (2 b+r_+))\\
&T(M,r_+)=-\frac{\mathcal{X}}{\mathcal{Y}}.
\end{split}
\end{equation}
Therefore, by using Eq. (\ref{19}) one can obtain,
\begin{equation}\label{29}
\begin{split}
&\mathcal{A}=a^8(2 b^2+2 b r_++r_+^2)+a^6 r_+ \big(16 b^3+2 b^2 (9 r_++2)+b r_+ (2 \Xi^2 M+8 r_++5)+r_+^2 (r_++2)\big)+a^4 r_+^2 \\&\bigg[48 b^4+4 b^3 (2 \Xi^2 M+16 r_++5)+4 b^2 r_+ (8 \Xi^2 M+7 r_++8)+b r_+^2 (20 \Xi^2 M+2 r_++19)\\&-r_+^2 \big(\Xi^2 (M-4 M r_+)+(r_+-4) r_+\big)\bigg]
+a^2 r_+^3 (2 b+r_+) \bigg[32 b^4+4 b^3 (9 r_++4)+2 b^2 r_+ (8 \Xi^2 M+5 r_++11) \\&+b r_+ \big(4 \Xi^2 M (r_++1)+r_+ (11-2 r_+)\big)-(r_+-2) r_+^3\bigg]\\
&+(2 b+1) r_+^4 (2 b+r_+)^2 \bigg[4 b^2 (b-\Xi^2 M)+r_+^2 (b-3 \Xi^2 M)+4 b r_+ (b-\Xi^2 M)\bigg]\\
&\mathcal{B}=\bigg[2 \pi r_+^2 (2 b + r_+)^2 (a^2 + r_+ + 2 b r_+)^2 (a^2 + r_+ (2 b + r_+))^2\bigg]\\
&\mathcal{C}=3 \bigg[a^4 (b+r_+)+a^2 r_+ \big(4 b^2+6 b r_++r_+ (2 r_+-\Xi^2 M)\big)+r_+^2 (2 b+r_+) \big(3 r_+ (b-\Xi^2 M)+2 b (b-\Xi^2 M)+r_+^2\big)\bigg]\\
&\mathcal{D}=4 \pi  r_+^2 (2 b+r_+)^2 \big(a^2+r_+ (2 b+r_+)\big)^2\\
&\mu =\frac{\mathcal{A}/\mathcal{B}}{\mathcal{C}/\mathcal{D}}.
\end{split}
\end{equation}
We express the mass $M$ and temperature $T$ in terms of the pressure P and the radius $r_+$ of a black hole, using different equations,
\begin{equation}\label{30}
M(P,r_+)=\frac{\left(a^2+r_+ (2 b+r_+)\right) (8 \pi  P r_+ (2 b+r_+)+3)}{6 \Xi^2r_+},
\end{equation}
\begin{equation}\label{31}
T(P,r_+)=\frac{a^2 \left(8 \pi  P r_+^2-3\right)+r_+^2 (8 \pi  P (2 b+r_+) (2 b+3 r_+)+3)}{12 \pi  r_+ \left(a^2+2 b r_++r_+\right)}.
\end{equation}
The $V$ for the AKS black hole is calculated as follows\cite{45},
\begin{equation}\label{32}
V=\frac{2\pi\big[a^2+r_+(2b+r_+)\big]\big[a^2\big[r_+(2b+r_+)+\ell^2\big]+2\Xi\ell^2r_+(b+r_+)\big]}{3\Xi^2\ell^2r_+}.
\end{equation}
So, by using the Eqs. (\ref{19}), (\ref{21}), (\ref{22}), and (\ref{23}) one can obtain,
\begin{equation}\label{33}
\begin{split}
&T_i=(2 b+r_+)(a^2+r_+ (2 b+r_+))\\&\times\bigg[a^4(8 \pi  P_i r_+^2+3)+a^2 r_+\bigg(32 \pi  b^2 P_i r_++4 b(32 \pi  P_i r_+^2+3)+3(24 \pi  P_i r_+^3+r_++2)\bigg)\\&+16 \pi  (2 b+1) P_i r_+^4 (4 b+3 r_+)\bigg]\\&\bigg/12 \pi  r_+^2(a^2+2 b r_++r_+)^2\big(a^2+(2 b+r_+) (2 b+3 r_+)\big).
\end{split}
\end{equation}
Also, with respect to Eq. (\ref{30}) we will have,
\begin{equation}\label{34}
\begin{split}
T_i=\frac{a^2 \left(8 \pi  r_+^2 P_i-3\right)+r_+^2 (8 \pi  P_i (2 b+r_+) (2 b+3 r_+)+3)}{12 \pi  r_+ \left(a^2+2 b r_++r_+\right)}.
\end{split}
\end{equation}
We have drawn the isenthalpic curves and the inversion curve of the AKS black hole for various free parameter values in each plot of Fig. (\ref{m5}). In every subfigure, three isenthalpic curves with different M are visible, along with the corresponding inversion curve that occurs at the highest point of the isenthalpic curves. We denote each isenthalpic curve's inversion temperature and pressure as $T_i$ and $P_i$, respectively. The isenthalpic curves are divided into two regions by the inversion curve: for $P < P_i$, the isenthalpic curve has a positive slope, indicating that the black hole undergoes cooling during the expansion process. With the changes in the parameters of the black hole, we found that the slope of the figures will change, and these changes are visible in each subfigure. Because an analytical solution is difficult to obtain, we used a numerical method to draw graphs. Assuming the parameter $b$ is zero, our equations and graphs will simplify to the Kerr–AdS black holes, whose results are thoroughly discussed in \cite{48}.
However, we can demonstrate the relation between the inversion temperature $T_i$ and the inversion pressure $P_i$ by using the equations given above. Figs. ((\ref{6a})-(\ref{6d})) displays the inversion curves for different free parameters of the AKS black holes. The inversion curve has only one branch. The inversion temperature increases gradually with the inversion pressure, but the slope of the inversion curves becomes smaller. We can also observe some fine structures in the cases of subfigures (\ref{m6}). Inversion curves change with the free parameters that are specified in each plot. It was very difficult to find the analytical solution for drawing the graphs, so we used numerical solutions to draw them. Moreover, we note that due to the huge difference in the values of the horizontal and vertical graphs corresponding to the change of the free parameters, we have drawn the figures separately.
In Fig. (\ref{m7}), we plot the Hawking temperature as a function of the horizon, which is the boundary of the black hole. The horizon can be affected by the presence of other fields or dimensions, which are represented by the free parameters in our model. We consider the free parameters constant for each plot, but we vary them across different plots to see how they influence the Hawking temperature. In each subfigure, we can observe zero points, where the Hawking temperature becomes zero for different free parameters. This means that the black hole became extremal at these points. These zero points correspond to the divergence points of the JTC, which is a quantity that describes the temperature change when a gas expands or compresses at constant enthalpy. In the study of Kerr black hole\cite{48}, it was shown that the rotation parameter "a" does not have much effect on the Joule-Thomson representation. For example, it is practically eliminated in the ratio of the $T_i^{min}/T_C$, and the value of this ratio is always a constant value of 1/2. In this paper, we found that the parameters related to AKS black hole, i.e. a and b, can play a vital role in the representation of the value of the $T_i^{min}/T_C$. The value of this ratio increases with the reduction of these parameters. For this purpose, due to the structural complexities of this black hole, we could not use the analytical method to obtain the critical points of the black hole, so we resorted to the approximate method and the numerical calculations. In the \cite{45}, the value of critical temperature and critical pressure has been calculated. We calculated some values for this black hole according to table (\ref{n4}). As evident, regarding various values of free parameters, the $T_i^{min}/T_C$ is obtained, which shows the obvious effect of parameters a and b. In the Kerr-AdS geometry, the rotation parameter has a minimal effect on the ratio $( T_{min}^{i} /T_c )$ because the thermodynamic properties are primarily governed by the AdS background and the black hole's mass. The rotation parameter, while influencing the black hole's angular momentum and ergosphere, does not significantly alter the overall thermodynamic balance between the inversion and critical temperatures. This indicates that the rotational effects are secondary to the dominant influence of the AdS space and the black hole's mass. The AdS background provides a stabilizing effect, ensuring that the primary thermodynamic properties remain largely unaffected by the rotation parameter. In the AKS black hole, the presence of additional fields such as the dilaton and axion fields introduces new interactions that modify the thermodynamic properties. These fields affect the black hole's mass and angular momentum differently compared to the Kerr-AdS case. As a result, the rotation parameter has a more pronounced effect on the ratio $( T_{min}^{i} /T_c )$. The altered behavior in the AKS case highlights the complex interplay between the black hole's rotation, the additional fields, and the AdS background, leading to a more intricate thermodynamic structure. The dilaton and axion fields introduce additional degrees of freedom and interactions that significantly impact the black hole's thermodynamics. These fields can alter the distribution of mass and angular momentum, leading to changes in the inversion temperature and the overall thermodynamic behavior. The rotation parameter, in this case, becomes more influential because it interacts with the dilaton and axion fields, creating a more complex and dynamic thermodynamic environment. This interplay results in a more nuanced understanding of the phase transitions and stability of the AKS black hole, as the additional fields and rotation parameters collectively shape the thermodynamic landscape.
\section{Conclusion}
In this article, we investigated the behavior of two structurally different black holes under the JTE thermodynamic process. The JTE is a process that cools or heats a system by changing its pressure and volume at constant enthalpy, and its main goal is to study the behavior of the $\mu$ coefficient, which describes the change in temperature during the expansion or compression of a system. But before presenting the results, we must first explain the motivations that led us to choose these particular black holes. Recently, an EPYM black hole with AdS structure under the JTE process was investigated in a research paper\cite{36,36.1}. This raised the question in our mind if fields in the Maxwellian form are added to the elements in the action of the above article, can this change and addition of a special form of the field cause a different thermodynamic behavior in their JT representation?
Similarly, in the past, the behavior of various forms of rotating black holes was also investigated in this thermodynamic process. But what effect can the addition of Sen-type dilaton fields and antisymmetric tensor fields have on its thermodynamic properties and JT representation? This was a question that could be a good motivation for our investigation. For this purpose, we plotted the isenthalpic curves and the inversion curves for each of the two black holes, for different values of the free parameters, which can be referred to the following results for each.
For the AEMPYM black hole, the Hawking temperature has zero points that depend on the free parameters and the horizon radius, and the JTC diverges at these points. The isenthalpic curves are divided into two regions by the inversion curve: for $P < P_i$, the isenthalpic curve has a positive slope, indicating that the black hole is cooling during the expansion process, and for $P > P_i$, the isenthalpic curve has a negative slope, which means that the black hole experiences heating during the expansion process. The inversion curves for different free parameters show that the inversion curve has only one branch and the inversion temperature increases steadily with the inversion pressure, but the slope of the inversion curves becomes smaller. Also, the graphs showed that for low pressure, the inversion temperature varies with the free parameters specified in each graph. Studying the effect of parameters was interesting for us because we observed that the slope of the inversion curve increases with changes in the different values of each parameter.
But if we want to discuss the effect of Maxwell's charges as a special result of our work, we must state that, we have found that the ratio of minimum to critical temperature for a 4-dimensional black hole with Yang-Mills hair depends on the values of $q$, $M$, and $C$. For $q=1/2$, the ratio is exactly equal to $1/2$ for all values of $C$. For $1/2<q<1$, the ratio approaches $1/2$ as $C$ increases. For $0<q<1/2$, the ratio deviates from $1/2$ as $C$ increases. These results show that the Maxwell charge can affect the thermodynamic behavior of the black hole and make it more or less similar to a 4-dimensional charged black hole. Also, for the AKS black hole, we found that its parameters, i.e. a and b, can play a vital role in the representation of the value of the ratio of $T_i^{min}/T_C$ so that by reducing these parameters, the value of the ratio increases. The results are shown in Fig. (\ref{m4}) and are summarized in tables (\ref{n1}) to (\ref{n5}).
For the AKS black, It was very difficult to find the analytical solution for drawing the graphs, so we used numerical solutions to draw them. For $P < P_i$, the isenthalpic curve has a positive slope, indicating that the black hole is cooling during the expansion process, but with the increases of the black hole parameters, we found that the slope of the figures will change.
Also, if we assume the parameter b ( dyonic charge) to be zero, our equations and graphs will simplify to the Kerr-AdS black holes, whose results are thoroughly discussed in\cite{48}.
In general, it can be said about this black hole the isenthalpic curves of the AKS black hole show cooling or heating behavior depending on the inversion curve, which is affected by the mass and the parameter $b, a$ of the black hole. The inversion curve has a single branch with a positive slope that decreases with the free parameters, and the inversion temperature and pressure are related by some equations that are numerically solved. The Hawking temperature of the AKS black hole has zero points that depend on the free parameters and the horizon radius and lead the JTC to reflect divergence behavior.
\begin{center}
\begin{table}[H]
  \centering
\begin{tabular}{|p{4cm}|p{4cm}||p{4cm}||p{4cm}|}
  \hline
  $q=0.3, n=2, Q=1$  & \hspace{1cm} $T_i^{min}$  & \hspace{1cm} $T_{c}$ & \hspace{1cm} $T_i^{min}/T_{c}$ \\[3mm]
   \hline
  \hspace{1cm} $C=0.1$ & \hspace{1cm} $0.171532$ & \hspace{1cm} $0.355378$ & \hspace{1cm} $0.482674$\\[3mm]
   \hline
  \hspace{1cm} $C=0.2$ & \hspace{1cm} $0.068796$ & \hspace{1cm} $0.148927$ &\hspace{1cm} $0.461947$  \\[3mm]
  \hline
  \hspace{1cm} $C=0.3$ & \hspace{1cm} $0.035473$ & \hspace{1cm} $0.082056$ & \hspace{1cm} $0.432299$ \\[3mm]
  \hline
  \hspace{1cm} $C=1$ & \hspace{1cm} Not Exist & \hspace{1cm} Not Exist & \hspace{1cm} Not Exist \\[3mm]
  \hline
  \hspace{1cm}  $C=5$ & \hspace{1cm} Not Exist & \hspace{1cm} Not Exist & \hspace{1cm} Not Exist  \\[3mm]
  \hline
  \hspace{1cm} $C=10$ & \hspace{1cm} Not Exist & \hspace{1cm} Not Exist & \hspace{1cm} Not Exist \\[3mm]
  \hline
\end{tabular}
\caption{Summary of the results for the AEMPYM black hole for $q=0.3$}\label{n1}
\end{table}
 \end{center}

\begin{center}
\begin{table}[H]
  \centering
\begin{tabular}{|p{4cm}|p{4cm}||p{4cm}||p{4cm}|}
  \hline
  $q=0.5, n=2, Q=1$  & \hspace{1cm} $T_i^{min}$  & \hspace{1cm} $T_{c}$ & \hspace{1cm} $T_i^{min}/T_{c}$ \\[3mm]
   \hline
  \hspace{1cm} $C=0.1$ & \hspace{1cm} $0.034331$ & \hspace{1cm} $0.068662$ & \hspace{1cm} $0.5$\\[3mm]
   \hline
  \hspace{1cm} $C=0.2$ & \hspace{1cm} $0.017165$ & \hspace{1cm} $0.034330$ &\hspace{1cm} $0.5$  \\[3mm]
  \hline
  \hspace{1cm} $C=0.3$ & \hspace{1cm} $0.011443$ & \hspace{1cm} $0.022886$ & \hspace{1cm} $0.5$ \\[3mm]
  \hline
  \hspace{1cm} $C=1$ & \hspace{1cm} $0.003433$ & \hspace{1cm} $0.006866$ & \hspace{1cm} $0.5$ \\[3mm]
  \hline
  \hspace{1cm}  $C=5$ & \hspace{1cm} $0.000686$ & \hspace{1cm} $0.001372$ & \hspace{1cm} $0.5$  \\[3mm]
  \hline
  \hspace{1cm} $C=10$ & \hspace{1cm} $0.000343$ & \hspace{1cm} $0.000686$ & \hspace{1cm} $0.5$ \\[3mm]
  \hline
\end{tabular}
\caption{Summary of the results for the AEMPYM black hole for $q=0.5$}\label{n2}
\end{table}
 \end{center}

\begin{center}
\begin{table}[H]
  \centering
\begin{tabular}{|p{4cm}|p{4cm}||p{4cm}||p{4cm}|}
  \hline
  $q=0.9, n=2, Q=1$  & \hspace{1cm} $T_i^{min}$  & \hspace{1cm} $T_{c}$ & \hspace{1cm} $T_i^{min}/T_{c}$ \\[3mm]
   \hline
  \hspace{1cm} $C=0.1$ & \hspace{1cm} $0.019146$ & \hspace{1cm} $0.03874$ & \hspace{1cm} $0.493657$\\[3mm]
   \hline
  \hspace{1cm} $C=0.2$ & \hspace{1cm} $0.018873$ & \hspace{1cm} $0.038208$ &\hspace{1cm} $0493962$  \\[3mm]
  \hline
  \hspace{1cm} $C=0.3$ & \hspace{1cm} $0.018446$ & \hspace{1cm} $0.037309$ & \hspace{1cm} $0.494405$ \\[3mm]
  \hline
  \hspace{1cm} $C=1$ & \hspace{1cm} $0.013846$ & \hspace{1cm} $0.027832$ & \hspace{1cm} $0497502$ \\[3mm]
  \hline
  \hspace{1cm}  $C=5$ & \hspace{1cm} $0.004141$ & \hspace{1cm} $0.008286$ & \hspace{1cm} $0.499762$  \\[3mm]
  \hline
  \hspace{1cm} $C=10$ & \hspace{1cm} $0.002133$ & \hspace{1cm} $0.004267$ & \hspace{1cm} $0.499921$ \\[3mm]
  \hline
\end{tabular}
\caption{Summary of the results for the AEMPYM black hole for $q=0.9$}\label{n3}
\end{table}
 \end{center}
\begin{center}
\begin{table}[H]
  \centering
\begin{tabular}{|p{4cm}|p{4cm}||p{4cm}||p{4cm}|}
  \hline
  \hspace{1cm} ($a , b$)  & \hspace{1cm} $T_i^{min}$  & \hspace{1cm} $T_{c}$ & \hspace{1cm} $T_i^{min}/T_{c}$ \\[3mm]
   \hline
  ($0.0099$, $0.00499$) & \hspace{1cm} $0.0787687$ & \hspace{1cm} $0.335030$ & \hspace{1cm} $0.235109$\\[3mm]
   \hline
  ($0.00972$, $0.00486$) & \hspace{1cm} $0.0782007$ & \hspace{1cm} $0.2302899$ &\hspace{1cm} $0.33957$  \\[3mm]
  \hline
  ($0.00951$, $0.00475$) & \hspace{1cm} $0.078821$ & \hspace{1cm} $0.156982$ & \hspace{1cm} $0.5$ \\[3mm]
  \hline
  ($0.00937$, $0.00468$) & \hspace{1cm} $0.0788340$ & \hspace{1cm} $0.116006$ & \hspace{1cm} $0.67379$ \\[3mm]
  \hline
  ($0.00892$, $0.00442$) & \hspace{1cm} $0.776375$ & \hspace{1cm} $0.0454247$ & \hspace{1cm} $1.709$  \\[3mm]
  \hline
\end{tabular}
\caption{Summary of the results for the AKS black hole}\label{n4}
\end{table}
 \end{center}

\begin{center}
\begin{table}[H]
  \centering
\begin{tabular}{|p{4.2cm}|p{4cm}||p{4cm}||p{4cm}|}
  \hline
  \hspace{0.7cm} AEMPYM BH  & \hspace{1cm} $q>1$, $C\gg1$  & \hspace{1cm} $T_{min}/T_{c}=\frac{1}{2}$ & \hspace{1cm} This paper \\[3mm]
   \hline
  \hspace{0.7cm} AEMPYM BH  & $q=0.5$,  all values of  $"C"$ & \hspace{1cm} $T_{min}/T_{c}\simeq\frac{1}{2}$ & \hspace{1cm} This paper \\[3mm]
   \hline
  \hspace{0.7cm} AEMPYM BH & \hspace{1cm}  $0.5<q\leq1$ , $C\gg1$ & \hspace{1cm}$T_{min}/T_{c}\simeq\frac{1}{2}$ & \hspace{1cm} This paper \\[3mm]
  \hline
  \hspace{0.7cm} AKS BH & \hspace{1cm} ($a=0.00951$, $b=0.00475$) & \hspace{1cm} $T_{min}/T_{c}\simeq\frac{1}{2}$ & \hspace{1cm} This paper \\[3mm]
  \hline
  \hline
  Van der Waals fluid & \hspace{1cm} Exist & \hspace{1cm} $T_{min}/T_{c}=\frac{3}{4}$ & \hspace{1cm} \cite{21} \\[3mm]
  \hline
  \hspace{1cm} RN-AdS BH & \hspace{1cm} Exist & \hspace{1cm} $T_{min}/T_{c}=\frac{1}{2}$ & \hspace{1cm} \cite{23} \\[3mm]
  \hline
  d-dimensional AdS BH & \hspace{1cm} Exist & \hspace{1cm} $T_{min}/T_{c}<\frac{1}{2}$ & \hspace{1cm} \cite{1002}  \\[3mm]
  \hline
  Gauss-Bonnet BH & \hspace{1cm} Exist & $T_{min}/T_{c}=0.4765$ & \hspace{1cm} \cite{1003} \\[3mm]
  \hline
  \hspace{1cm} torus-like BH & \hspace{1cm} Not Exist & $T_{min}/T_{c}$= Not Exist & \hspace{1cm} \cite{1004}  \\[3mm]
  \hline
  \hspace{1cm} BTZ BH & \hspace{1cm} Not Exist & $T_{min}/T_{c}$= Not Exist & \hspace{1cm} \cite{1005} \\[3mm]
  \hline
\end{tabular}
\caption{Summary of the results for the AEMPYM and AKS black holes compare with other results}\label{n5}
\end{table}
 \end{center}

\section{Appendix: Numerical technique}
To construct the temperature-pressure graph $( T(M, P) )$ as depicted in Figure (2), it is essential to first determine the radius of the event horizon $(r_+ )$ as a function of pressure $( P )$, utilizing Eq.(\ref{16}). Subsequently, by substituting this relationship into Eq.(17), we can generate the desired graph. However, the intricate nature of Eq.(\ref{16}) precludes us from deriving an analytical expression for the event horizon radius $( r_+ )$ in terms of pressure $( P )$. As a result, we resort to numerical methods for this purpose. By holding the mass $( M )$, the charge $( Q )$, and the constant $( C )$ fixed, and by inputting various radii for the event horizon into Eqs. (\ref{16}) and (\ref{17}), we can calculate the corresponding temperatures and pressures for different events. This numerical approach applies to any scenario that defies analytical solutions. Indeed, this method can be employed for both types of black holes as delineated by the equations presented in the text. To illustrate this numerical technique, we provide an example below, which corresponds to the scenario outlined in Fig. (\ref{m2}). Please note that the numerical approach for Fig. (\ref{m2}) will be placed in the following table (\ref{n6}) for the AEMPYM black hole with respect to its free parameters demonstrating the numerical method to derive the temperature and pressure values associated with the event horizon of the black hole. Also, to draw Fig. (\ref{m5}) for the AKS black hole, a similar process can be done according to Eqs. (\ref{26}) and (\ref{30}), as well as its free parameters in the text.
\begin{center}
\begin{table}[H]
  \centering
\begin{tabular}{|p{5.5cm}|p{5.5cm}|}
  \hline
  \hspace{2.5cm} P  & \hspace{2.5cm} T  \\[2mm]
   \hline
 \hspace{2cm}0.196046 & \hspace{2cm}0.011344 \\[2mm]
  \hline
  \hspace{2cm}0.193908&\hspace{2cm}0.0822105  \\[2mm]
  \hline
  \hspace{2cm}0.189191&\hspace{2cm}0.137814  \\[2mm]
  \hline
  \hspace{2cm}0.182693&\hspace{2cm}0.181027  \\[2mm]
  \hline
  \hspace{2cm}0.17897&\hspace{2cm} 0.198729   \\[2mm]
   \hline
  \hspace{2cm}0.170874&\hspace{2cm}0.227563\\[2mm]
      \hline
  \hspace{2cm}0.162223&\hspace{2cm}0.249009 \\[2mm]
      \hline
  \hspace{2cm}0.153304&\hspace{2cm}0.264447  \\[2mm]
      \hline
  \hspace{2cm}0.144326&\hspace{2cm}0.275003\\[2mm]
      \hline
  \hspace{2cm}0.135443&\hspace{2cm}0.281598\\[2mm]
     \hline
  \hspace{2cm}0.131072&\hspace{2cm}0.283651\\[2mm]
      \hline
  \hspace{2cm}0.122525&\hspace{2cm}0.285673\\[2mm]
      \hline
 \hspace{2cm}0.119611&\hspace{2cm}0.285807\\[2mm]
      \hline
 \hspace{2cm}0.118364&\hspace{2cm}0.285783\\[2mm]
      \hline
 \hspace{2cm}0.106396&\hspace{2cm}0.283216\\[2mm]
     \hline
 \hspace{2cm}0.0883264&\hspace{2cm}0.272013\\[2mm]
      \hline
  \hspace{2cm}0.0592559&\hspace{2cm}0.236967 \\[2mm]
     \hline
  \hspace{2cm}0.0382974&\hspace{2cm}0.198038\\[2mm]
      \hline
  \hspace{2cm}0.0234604&\hspace{2cm}0.161811\\[2mm]
      \hline
  \hspace{2cm}0.0130321&\hspace{2cm}0.130287\\[2mm]
      \hline
  \hspace{2cm}0.00572774&\hspace{2cm}0.103673 \\[2mm]
     \hline
  \hspace{2cm}0.000628043&\hspace{2cm}0.0815221\\[2mm]
      \hline
  \hspace{2cm}0.000212542&\hspace{2cm}0.0795286\\[2mm]
  \hline
\end{tabular}
\caption{$ M=4, q=1.5, n=3, C=0.5, Q=1$}\label{n6}
\end{table}
 \end{center}

\begin{figure}[H]
 \begin{center}
 \subfigure[]{
 \includegraphics[height=6.5cm,width=6.5cm]{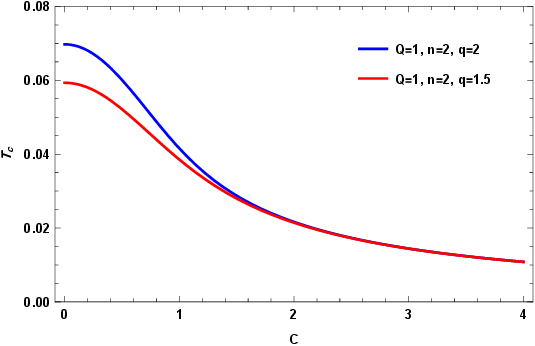}
 \label{8a}}
  \caption{\small{The plot of $T_c-C$. The solid blue line illustrated with $q=2, n=2$, and $Q=1$ and solid red line denoted with $q=1.5, n=2$, and $Q=1$ for the AEMPYM black hole}}
 \label{m8}
 \end{center}
 \end{figure}

To clarify the relationship between the critical temperature, $( T_c )$, and Maxwell's charge, $( C )$, we analyzed the (AEMPYM) black holes for different values $Q=1, n=2, q=2$ and $Q=1, n=2, q=1.5$. Using the critical condition, we first obtain $r_{+,c}$. Then by placing it in the critical temperature, we calculate the direct relationship between the critical temperature and the Maxwell charge. Also, to better understand this relationship, we draw Fig. (\ref{m8}). As shown in Fig. (\ref{m8}), we find for larger values of $( C )$, the results tend to converge, becoming consistent across different $( q )$ values. Conversely, for Maxwell charge values less than 1, a higher critical temperature is noted for larger $( q )$ values.
\begin{equation}\label{34}
\begin{split}
r_{+,c}=\sqrt{2^{2/3} \sqrt[3]{2 c^6+\sqrt{7} \sqrt{4 c^6+7}+7}+2 c^2+\frac{2 \sqrt[3]{2} c^4}{\sqrt[3]{2 c^6+\sqrt{7} \sqrt{4 c^6+7}+7}}}.
\end{split}
\end{equation}
where for $Q=1, n=2, q=2$, we will have,
\begin{equation}\label{34}
\begin{split}
r_{+,c}=0.5 \sqrt{2 \sqrt{36 c^4+84.8528}+12 c^2}
\end{split}
\end{equation}
Then, we can obtain,
\begin{equation}\label{34}
\begin{split}
&\mathcal{X}_1=\bigg[-2 (2^{2/3} c \sqrt[3]{2 c^6+\sqrt{7} \sqrt{4 c^6+7}+7}+2 c^3+\frac{2 c^5}{\sqrt[3]{c^6+\frac{1}{2} (\sqrt{7} \sqrt{4 c^6+7}+7)}})^2\\&+(2^{2/3} \sqrt[3]{2 c^6+\sqrt{7} \sqrt{4 c^6+7}+7}+2 c^2+\frac{2 c^4}{\sqrt[3]{c^6+\frac{1}{2} (\sqrt{7} \sqrt{4 c^6+7}+7)}})^3-8\bigg]\\
&\mathcal{X}_2=\bigg[2 \pi  (2^{2/3} \sqrt[3]{2 c^6+\sqrt{7} \sqrt{4 c^6+7}+7}+2 c^2+\frac{2 c^4}{\sqrt[3]{c^6+\frac{1}{2} (\sqrt{7} \sqrt{4 c^6+7}+7)}})^{7/2}\bigg]\\
&T_c=\frac{\mathcal{X}_1}{\mathcal{X}_2}
\end{split}
\end{equation}
Also, for $Q=1, n=2, q=1.5$, we will have,
\begin{equation}\label{34}
\begin{split}
r_{+,c}=0.5 \sqrt{2 \sqrt{36 c^4+84.8528}+12 c^2}
\end{split}
\end{equation}
In this case, the corresponding critical temperature is obtained from the following equation,
\begin{equation}\label{34}
\begin{split}
&\mathcal{Y}_1=162.975\bigg[0.00195313 \bigg(2 \sqrt{36 c^4+84.8528}+12 c^2\bigg)^{5}-0.015625 c^2 \bigg(2 \sqrt{36 c^4+84.8528}+12 c^2\bigg)^{4}\\&-1.06066 \bigg(1 \sqrt{36 c^4+84.8528}+6 c^2\bigg)^3\bigg]\\
&\mathcal{Y}_2=(2 \sqrt{36 c^4+84.8528}+12 c^2)^{5.5}\\
&T_c=\frac{\mathcal{Y}_1}{\mathcal{Y}_2}
\end{split}
\end{equation}

\end{document}